\documentclass[%
 reprint,
superscriptaddress,
 amsmath,amssymb,
 aps,
prb,
showkeys
]{revtex4-2}

\usepackage{graphicx}
\usepackage{dcolumn}
\usepackage{bm}
\usepackage[utf8]{inputenc}

\begin{document}

\title {Polyhedral distortions and unusual magnetic order in spinel FeMn$_{2}$O$_{4}$}
\altaffiliation{
Copyright  notice: This  manuscript  has  been  authored  by  UT-Battelle, LLC under Contract No. DE-AC05-00OR22725 with the U.S.  Department  of  Energy.   
The  United  States  Government  retains  and  the  publisher,  by  accepting  the  article  for  publication, 
acknowledges  that  the  United  States  Government  retains  a  non-exclusive, paid-up, irrevocable, world-wide license to publish or reproduce the published form of this manuscript, 
or allow others to do so, for United States Government purposes.  
The Department of Energy will provide public access to these results of federally sponsored  research  in  accordance  with  the  DOE  Public  Access  Plan 
(http://energy.gov/downloads/doe-public-access-plan)}

\author{Qiang Zhang}
 \email{zhangq6@ornl.gov}

 \affiliation{Department of Physics and Astronomy, Louisiana State University, Baton Rouge, Louisiana 70803, USA}
\affiliation{Neutron Scattering Division, Oak Ridge National Laboratory, Oak Ridge, Tennessee 37831, USA}

\author{Wei Tian}
  \affiliation{Neutron Scattering Division, Oak Ridge National Laboratory, Oak Ridge, Tennessee 37831, USA}
  
  \author{Roshan K. Nepal}
 \affiliation{Department of Physics and Astronomy, Louisiana State University, Baton Rouge, Louisiana 70803, USA}

\author{Ashfia Huq}\altaffiliation{Current address: Sandia National Laboratories,
Livermore, CA 94551, USA}
   \affiliation{Neutron Scattering Division, Oak Ridge National Laboratory, Oak Ridge, Tennessee 37831, USA}

 \author{Stephen Nagler}
  \affiliation{Neutron Scattering Division, Oak Ridge National Laboratory, Oak Ridge, Tennessee 37831, USA}

  \author{J. F. DiTusa}
\affiliation{Department of Physics and Astronomy, Louisiana State University, Baton Rouge, Louisiana 70803, USA}
\affiliation{Department of Physics, Indiana University-Purdue University Indianapolis, Indianapolis, IN 46202}
 
\author{Rongying Jin}
 \email{rjin@mailbox.sc.edu}
\affiliation{Department of Physics and Astronomy, Louisiana State University, Baton Rouge, Louisiana 70803, USA}
\affiliation{Center for Experimental Nanoscale Physics, Department of Physics and Astronomy, University of South Carolina, Columbia, SC 29208, USA}

\date{\today}

\begin{abstract}
Spinel compounds AB$_{2}$X$_{4}$ consist of both tetrahedral (AX$_{4}$) and octahedral (BX$_{6}$) environments with the former forming a diamond lattice and the latter a geometrically frustrated pyrochlore lattice. Exploring the fascinating physical properties and their correlations with structural features is critical in understanding these materials. FeMn$_{2}$O$_{4}$ has been reported to exhibit one structural transition and two successive magnetic transitions. Here, we report the polyhedral distortions and their correlations to the structural and two magnetic transitions in FeMn$_{2}$O$_{4}$ by employing the high-resolution neutron powder diffraction. The cation distribution is found to be ($Mn^{2+}_{0.9}Fe^{3+}_{0.1}$)$_{A}$($Mn^{3+}Fe^{3+}_{0.9}Mn^{2+}_{0.1}$)$_{B}$O$_{4}$. While large trigonal distortion is found even in the high-temperature cubic phase, the first-order cubic-tetragonal structural transition associated with the elongation of both tetrahedra and octahedra with shared oxygen atoms along the $c$ axis occurs at $T_{S} \approx$ 750 K, driven by the Jahn-Teller effect of the orbital active B-site Mn$^{3+}$ cation. Strong magnetoelastic coupling is unveiled at  $T_{N1}\approx 400$ K as manifested by the appearance of $N\acute{e}el$-type collinear ferrimagnetic order, an anomaly in both tetrahedral and octahedral distortions, as well as an anomalous decrease of the lattice constants $c$ and a weak anomaly of $a$. Upon cooling to $T_{N2}\approx65$ K, it evolves to a noncollinear ferrimagnetic order accompanied with the different moments at the split magnetic sites B1 and B2.  Only one-half of the B-site $Mn^{3+}$/$Fe^{3+}$ spins, i.e., the B2-site spins in the pyrochlore lattice are canted, which is a unique magnetic order among spinels. The canting angle between A-site and B2-site moments is $\sim$25$^{\rm{o}}$, but the B1-site moment stays antiparallel to the A-site moment even at 10 K. This noncollinear order is accompanied by a modification of the O-B-O bond angles in the octahedra without significant change in lattice constants or tetrahedral/octahedral distortion parameters, indicating a distinct magnetoelastic coupling. We demonstrate distinct roles of the A-site and B-site magnetic cations in the structural and magnetic properties of FeMn$_{2}$O$_{4}$. Our study indicates that FeMn$_{2}$O$_{4}$ is a wonderful platform to unveil interesting magnetic order and to investigate their correlations with polyhedral distortions and lattice.

\end{abstract}

\maketitle
\section{\label{sec:level1}INTRODUCTION}

Spinel compounds AB$_{2}$X$_{4}$ (X=O, S, Se or Te) are known to exhibit a variety of interesting physical properties\cite{Tsurkan2021} such as multiferroicity \cite{Katsura2005,Giovannetti2011}, high electrochemical activity as cathode material for Li-ion batteries\cite{Song2012,Feng2012,Lee2014}, colossal magnetoresistivity\cite{Ramirez1997}, topological semimetallicity\cite{Xu2011}, etc. These properties are largely related to their unique crystal structure consisting of both AX$_{4}$ tetrahedra and BX$_{6}$ octahedra. As illustrated in Fig.1 (a), the A-site cations form a diamond lattice, while the B-site cations form a pyrochlore lattice which is geometrically frustrated.\cite{Tsurkan2021} The interplay between lattice, orbital and spin degrees of freedom is key in understanding the physical properties of this materials family. The system is even more interesting when both A and B sites are occupied by magnetic cations in spinel oxides, for instance MnV$_{2}$O$_{4}$\cite{Garlea2008} and FeV$_{2}$O$_{4}$ \cite{Zhang2012,MacDougall2012}. In these materials, a $N\acute{e}el$-type collinear ferrimagnetic (CFI) order is frequently observed at higher temperatures, but transformed to a noncollinear ferrimagnetic (NCFI) order at lower temperatures. 
A common ground-state  magnetic structure is a Yafet-Kittel type FI order where the A-site and B-site spins form triangular arrangements \cite{YAFET1952,Garlea2008,MacDougall2012} due to the competing magnetic interactions. For this magnetic structure, all the B-site spins in the pyrochlore lattice are canted relative to the A-site moment direction. The net moment at the B site is antiparallel to that at the A site, leading to uncompensated moments. It is of great interest to explore novel or unusual magnetic states and to investigate the correlation between magnetic order and tetrahedral/octahedral distortions in spinel oxides.

FeMn$_{2}$O$_{4}$ is one of the spinel oxides involving distinct magnetic cations at the A and B sites.\cite{BOUCHER1969,BRABERS1971} 
 It has been well documented from
 M\"{o}ssbauer\cite{Tanaka1963}, x-ray absorption spectroscopy and magnetic circular dichroism measurements\cite{Lee2008}that Mn$^{2+}$ cations mainly occupy the A site, whereas Mn$^{3+}$ and Fe$^{3+} $ cations share the B site. In this sense, FeMn$_{2}$O$_{4}$ should be written as Mn$_{A}$(MnFe)$_{B}$O$_{4}$\cite{Tanaka1963,Lee2008}. FeMn$_{2}$O$_{4}$ shows a cubic-to-tetragonal structural transition temperatures between 520 and 623 K with a strong dependence on the stoichiometry, cation distribution, or the degree of inversion\cite{BRABERS1971,Nepal2018}. Our previous neutron diffraction work at high temperatures detected a peak splitting showing a structural transition at $\sim$ 600 K.\cite{Nepal2018} On the other hand, the two magnetic transition temperatures have been reported to be less sensitive to the sample  preparation conditions, with  $T_{N1}\sim$ 373-400 K and $T_{N2}\sim$ 50-75 K, respectively\cite{BOUCHER1969,Nepal2018}. According to an early neutron diffraction experiment\cite{BOUCHER1969}, a collinear or noncollinear magnetic order can occur in $T_{N2}<T<T_{N1}$ but it must be a noncollinear order below $T_{N2}$. The proposed ground-state magnetic order was very complicated with a low-symmetry magnetic space group $C2'/m'$ derived from a monoclinic space group $C2/m$ with the following characteristics: (1). The A-site moment does not point along any high symmetry axis, instead forming an angle of 172$^{\rm o}$ to the [100]$_{T}$ direction in the tetragonal notation; (2). The B site is divided into 4 magnetic sublattices with different canting angles. Given that the crystal structure is tetragonal with a much higher symmetry $I4_{1}/amd$\cite{BOUCHER1969,BRABERS1971,Nepal2018} than this magnetic space group, this leaves the possibility of considering other more symmetric magnetic ground-state structures. In addition, it remains an open yet very important question in FeMn$_{2}$O$_{4}$ if there exists a correlation between magnetism, lattice and tetrahedral/octahedral distortions.

Here, we report the crystal and magnetic structures, and  their correlations with particular attention to polyhedral distortions as determined by the high-resolution neutron diffraction technique with a large \textit{Q} range. We find that there is large trigonal distortion persisting even within the high-temperature cubic phase. When cooling down, a  first-order cubic-to-tetragonal transition at $T_{S}$ is induced by the elongation of both tetrahedra and octahedra along the $c_{T}$ axis.   Two types of magnetoelastic couplings are found at these two magnetic transitions. A collinear FI order with the moments along the $a_{T}$ axis is found in $T_{N2} <T<T_{N1}$, which is accompanied with an anomalous reduction of the tetrahedral elongation, an enhanced octahedral distortion, and anomalies in the lattice constants. All these results indicate the existence of strong magnetoelastic coupling at $T_{N1}$. With further cooling to $T_{N2}$, the magnetic structure evolves to an unusual noncollinear FI order, where only half of the B-site spins are canted. The magnetic space group is $Imm\prime$a$\prime$ involving only two magnetic sublattices at the B site. This magnetic transformation at $T_{N2}$ induces anomalies in the O-B-O bond angles in octahedra. We demonstrate that the A-site Mn$^{2+}$, B-site Fe$^{3+}$ and Mn$^{3+}$ magnetic cations play distinct roles in the structural and magnetic phase transitions in FeMn$_{2}$O$_{4}$. We further shed light on the origin of the structural and magnetic ordering processes and the connection to competing interactions.

\section{EXPERIMENTAL DETAILS}

 High quality FeMn$_{2}$O$_{4}$ powder of 0.4 g was obtained by pulverizing the single crystals reported previously \cite{Nepal2018} for high-resolution neutron diffraction experiments at the time-of-flight neutron diffractometer POWGEN in Spallation Neutron Source (SNS), located at Oak Ridge National Laboratory (ORNL). A POWGEN sample changer (PAC) and a vacuum MICAS furnace were used to cover the temperature regions of 10-300 K and 300-1473 K, respectively. Helium exchange gas was sealed in the container for measurements in PAC, whereas the sample was placed in vacuum for the high-temperature measurements in MICAS furnace. The neutron bank with the center wavelength of 0.8 \AA{} was used to cover a wide \textit{Q} region of 0.9-11.8 \AA{}$^{-1}$.  Due to thermal hysteresis by the high-temperature structural transition, all low-temperature experiments were performed prior to high-temperature measurements. 
 
 To track the temperature dependence of the integrated intensity and linewidth of the low-\textit{Q} nuclear or magnetic peaks, we conducted a second experiment at the Fixed-Incident-Energy Triple-Axis Spectrometer HB1A in High Flux Isotope Reactor (HFIR) at ORNL, taking the advantage of the simple Gaussian peak profile function. The polycrystalline powder from the same batch of sample in our previous report\cite{Nepal2018} was used for the experiment at HB1A. A constant-wavelength neutron beam with $\lambda$= 2.36 \AA{} was used for data collection. A cryofurnace (JANIS) was employed to cover the temperature region between 5 K and 450 K. The helium exchange gas was sealed in the container for the measurements. The magnetic transition temperatures of these two samples are well consistent based on these two neutron experiments. Rietveld refinement of the neutron data was performed by the FullProf package\cite{Carvajal1993}. The symmetry-allowed magnetic structures were analyzed via the Bilbao Crystallographic Server\cite{Ivantchev:ks0038} and ISODISTORT\cite{Campbell2006}. To facilitate discussion, the cubic notation will be used to index the nuclear and magnetic peaks as well as bond length directions unless otherwise noted.

  \section{RESULTS}
 
\subsection{\label{sec:level2} First-order nature of the structural transition at $T_{S}$ }
 
\begin{figure}
\centering
\includegraphics[width=1\linewidth]{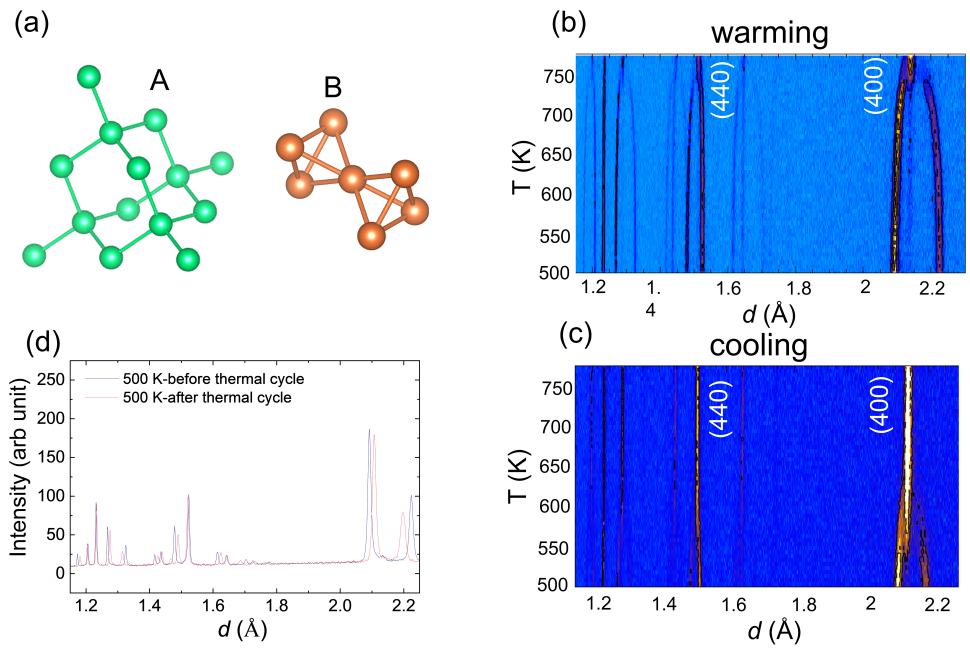}
\caption{(a). The Diamond and pyrochlore lattices consisting of the A-site and B-site ions, respectively. Temperature versus \textit{d} spacing of neutron diffraction contour plots via (b) warming and (c) cooling processes with a ramping rate of 1.5 K/min. (d) Comparison of the neutron diffraction patterns at 500 K before and after thermal cycle. }
\label{fig:1}
\end{figure}

 To identify the structural transition temperature $T_{S}$ and reveal the nature of the transition, neutron diffraction patterns were collected upon warming from 500 K to 773 K, followed by a cooling process at the ramping rate of 1.5 K/min for both warming and cooling. Neutron diffraction contour plots (temperature versus \textit{d} spacing) are displayed in Fig.~\ref{fig:1}(b)
 and (c) for warming and cooling processes, respectively. Clear peak splitting occurs on many nuclear Bragg peak positions, such as (400) and (440), confirming the existence of the cubic-tetragonal structural transition reported previously\cite{BOUCHER1969,BRABERS1971,Nepal2018}. Notably, there is large thermal hysteresis for the structural transition with $T_{S}\approx$ 750 K upon warming and $\approx$ 625 K via cooling, revealing the first-order nature. Previously, the structural transition was found to complete at $\approx$ 618 K\cite{Nepal2018}. The transition temperature is comparable with that upon the cooling process but lower than that during the warming process here. Note that the polycrystalline powder in Ref. 17 was synthesized using solid-state reaction method. The sample used at POWGEN was obtained from crashing the single crystals grown by the floating zone technology\cite{Nepal2018}. The different preparation conditions can affect the structural transition temperature as reported previously.\cite{BRABERS1971,BRABERS1971,Nepal2018}. On the other hand, these two experiments were conducted with different thermal histories. The data in Ref. 17 was collected with warming during the 2nd thermal cycling. Data shown in Fig. 1(b-c) was collected through the 1st thermal cycling. As displayed in Fig. 1(d), the neutron diffraction patterns taken at 500 K with one on warming and another on cooling show clear difference, for instance the Bragg peak positions and peak intensities. This indicates that the crystal structure is not completely restored at 500 K after the 1st thermal cycling. It is expected that the structural transition temperature with warming for the 2nd thermal cycling as done in previous report \cite{Nepal2018} would be different from that detected during the 1st thermal cycling here. 
   
  \begin{figure}
\centering
\includegraphics[width=1\linewidth]{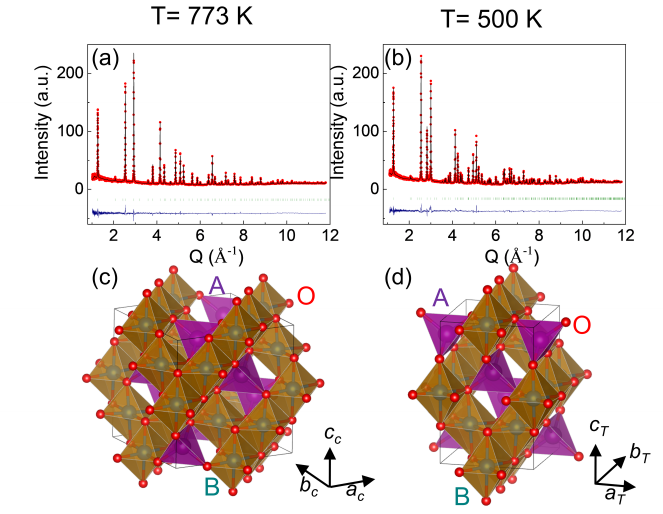}
\caption{(a-b). Rietveld refinement fits to high-resolution neutron diffraction patterns at (a) 773 K and (b) 500 K before thermal cycle, and (c-d) the crystal structures at (c) 773 K and (d) 500 K. }
\label{fig:2}
\end{figure}

\begin{table*} 
\centering
\setlength{\abovecaptionskip}{10pt}%
\setlength{\belowcaptionskip}{10pt}%
\caption{Refined atomic positions, isotropic temperature factors, occupancy, O-A-O bond angles in AO$_{4}$, O-B-O bond angles in BO$_{6}$, the tetrahedral distortion parameter D$_{\rm T}$ and the octahedral distortion parameter D$_{\rm O}$ from modeling high-resolution powder neutron diffraction data of FeMn$_{2}$O$_{4}$ at 773 K. Analysis of the nuclear Bragg reflections lead to the space group
selection of  $F d -3 m$ (No. 227) and indexed unit cell constants of $a =$ 8.562(3)  \AA{},  }
\renewcommand{\arraystretch}{0.8}
   \scalebox{0.8}{
\begin{tabular}{c|c|c|c|c|c|c|c|c|c|c}
\hline\hline
 atom & site&  x & y &  z& $B$ &  occupancy&  $\angle$ O-A-O($^{\rm o}$) & $\angle$ O-B-O ($^{\rm o}$) & D$_{\rm T}$ &   D$_{\rm O}$   \\
 \hline
Mn & 8a&    0.125&  0.125     &   0.125  & 1.675(5)& 0.897(6)&  & & & \\
Fe& 8a &    0.125&  0.125     &   0.125   &  1.675(5) &  0.103(6)&    & &  & \\ 
Fe & 16d &    0.5&  0.5     &   0.5  &   1.602(6)  & 0.448(5)&  & & & \\
Mn & 16d&    0.5&  0.5     &   0.5  &  1.602(6) &  0.552(5) & & & &  \\
O & 32e &    0.262(4)&  0.262(4)     &  0.262(4)  &   1.870(9) & 1.00(1)&   && & \\
 &   &   &       &  &    & &  109.47  & 96.34, 83.66&1 & 1\\
 \hline\hline  

\end{tabular}
 }
\label{crystal_1}
\end{table*}

 To further investigate these two crystal structures below and above the structural transition, high-resolution neutron diffraction patterns at 773 K and 500 K before thermal cycle are displayed in Fig.~\ref{fig:2}(a-b), respectively. As the inversion degree in some spinels such as MnFe$_{2}$O$_{4}$\cite{LEVY201515,Yang2009} is strongly dependent on the preparation conditions, particle size, and temperatures, it is of importance to determine the inversion at different temperatures in FeMn$_{2}$O$_{4}$. Rietveld analysis of neutron data over a large $Q$ coverage (up to $\approx$ 11.8 \AA{}$^{-1}$) allows us to obtain accurate the thermal parameters, site occupancy and inversion between the A and B sites. At 773 K, we find that FeMn$_{2}$O$_{4}$ crystallizes in the cubic structure with space group $F d -3 m$, as illustrated in the Fig. ~\ref{fig:2}(c). We confirm previous reports\cite{Tanaka1963,Lee2008} showing that the Mn$^{2+}$ ions occupy the A site, whereas mixed Mn$^{3+}$ and Fe$^{3+}$ ions occupy the B site randomly without evidence for cation ordering. Considering the difference in the neutron scattering lengths of Mn and Fe, we can determine that there is $\approx$ 10 $\%$ of Fe$^{3+}$ on the A site, with corresponding $\approx$ 10 $\%$ Mn$^{2+}$ on the B site in our sample. There is no appreciable change in the inversion within uncertainty between 500 K and 773 K. Thus, the cation distribution of our sample should be written as  
($Mn^{2+}_{0.9}Fe^{3+}_{0.1}$)$_{A}$($Mn^{3+}Fe^{3+}_{0.9}Mn^{2+}_{0.1}$)$_{B}$O$_{4}$. At 500 K (below $T_{S}$), Rietveld analysis of neutron data shown in Fig.~\ref{fig:2} (b) confirms the tetragonal structure with the space group $I4_{1}/amd$ with a clear tetragonal distortion (see Fig.~\ref{fig:2}(d)), consistent with the previous reports\cite{BOUCHER1969,BRABERS1971}. The refined atomic positions, lattice constants, site occupancy and thermal parameters at 773 K and 500 K are summarized in Table ~\ref{crystal_1} and ~\ref{crystal_2}, respectively. 

 For comparison, neutron diffraction patterns collected at 500 K after thermal cycle are also refined as shown in Fig. S1 and the structural parameters are summarized in   
Table ~\ref{crystal_3}. The structure is still tetragonal with the same space group. Compared with the structural parameters at 500 K before thermal cycle, there are no appreciable changes in the atomic positions or site occupancy, including the inversion between the A and B sites.  The main differences are the lattice constants with slightly different thermal parameters. The lattice constant $a$ increases but $c$ decreases after thermal cycle, consistent with the reduced structural transition temperature during the cooling process.

\parshape=0

\begin{table*} 
\centering
\setlength{\abovecaptionskip}{10pt}
\setlength{\belowcaptionskip}{10pt}%
\caption{Refined atomic positions, isotropic temperature factors, occupancy, O-A-O bond angles in AO$_{4}$, O-B-O bond angles in BO$_{6}$, the tetrahedral distortion parameter D$_{\rm T}$ and the octahedral distortion parameter D$_{\rm O}$ from modeling high-resolution powder neutron diffraction data of FeMn$_{2}$O$_{4}$ at 500 K  prior to thermal cycle. Analysis of the nuclear Bragg reflections leads to the space group
selection of   $I4_{1}/amd$ (No. 141) with the unit cell constants of $a_{T} =$ 5.921(4)\AA{} and $c_{T} =$ 8.900(3)\AA{}. }
\renewcommand{\arraystretch}{0.8}
\begin{tabular}{c|c|c|c|c|c|c|c|c|c|c}
\hline\hline
 atom & site&  x & y &  z& $B$ &  occupancy&  $\angle$ O-A-O($^{\rm o}$) & $\angle$ O-B-O ($^{\rm o}$) & D$_{\rm T}$ &   D$_{\rm O}$   \\
 \hline
Mn &  4a&    0&  0.25     &   0.875  & 1.212(5)& 0.905(8)&  & & & \\
Fe&  4a &    0&  0.25     &   0.875   &  1.212(5) &  0.095(8)&    & &  & \\ 
Fe & 8d &    0&  0.5     &   0.5  &   0.952(7)  & 0.452(4)&  & & & \\
Mn & 8d&    0&  0.5     &   0.5   &  0.952(7) &  0.548(4) & & & &  \\
O & 16e &   0& 0.474(6)     &  0.261(4)  &  1.245(8) & 1.00(1)&   && & \\
 &   &   &       &  &    & &  110.88,106.69  & 96.35, 83.65, 96.01,83.99&1.027 & 1.068\\
 \hline\hline  

\end{tabular}
\label{crystal_2}
\end{table*}

\begin{table*} 
\centering
\setlength{\abovecaptionskip}{10pt}
\setlength{\belowcaptionskip}{10pt}%
\caption{Refined atomic positions, isotropic temperature factors, occupancy, O-A-O bond angles in AO$_{4}$, O-B-O bond angles in BO$_{6}$, the tetrahedral distortion parameter D$_{\rm T}$ and the octahedral distortion parameter D$_{\rm O}$ from modeling high-resolution powder neutron diffraction data of FeMn$_{2}$O$_{4}$ at 500 K after thermal cycle from 500 to 773 K and back to 500 K. Analysis of the nuclear Bragg reflections leads to the space group
selection of   $I4_{1}/amd$ (No. 141) and indexed unit cell constants of $a_{T} =$ 5.957(4)\AA{}, $c_{T} =$ 8.785(3)\AA{} }
\renewcommand{\arraystretch}{0.8}
\begin{tabular}{c|c|c|c|c|c|c|c|c|c|c}
\hline\hline
 atom & site&  x & y &  z& $B$ &  occupancy&  $\angle$ O-A-O($^{\rm o}$) & $\angle$ O-B-O ($^{\rm o}$) & D$_{\rm T}$ &   D$_{\rm O}$   \\
 \hline
Mn &  4a&    0&  0.25     &   0.875  & 1.163(6)& 0.897(10)&  & & & \\
Fe&  4a &    0&  0.25     &   0.875   &  1.163(6) &  0.103(10)&    & &  & \\ 
Fe & 8d &    0&  0.5     &   0.5  &   1.096(8)  & 0.452(5)&  & & & \\
Mn & 8d&    0&  0.5     &   0.5   &  1.096(8) &  0.548(5) & & & &  \\
O & 16e &   0& 0.475(3)     &  0.262(4)  &   1.374(5) & 1.02(4)&   && & \\
 &   &   &       &  &    & &  110.54,107.35  & 96.22, 83.78, 96.12,83.88&1.020 & 1.0443\\
 \hline\hline  

\end{tabular}
\label{crystal_3}
\end{table*}

\subsection{\label{sec:level2} Intrinsic trigonal distortion of the BO$_{6}$ octahedron}

\begin{figure}[ht]
\centering
\includegraphics[width=1\linewidth]{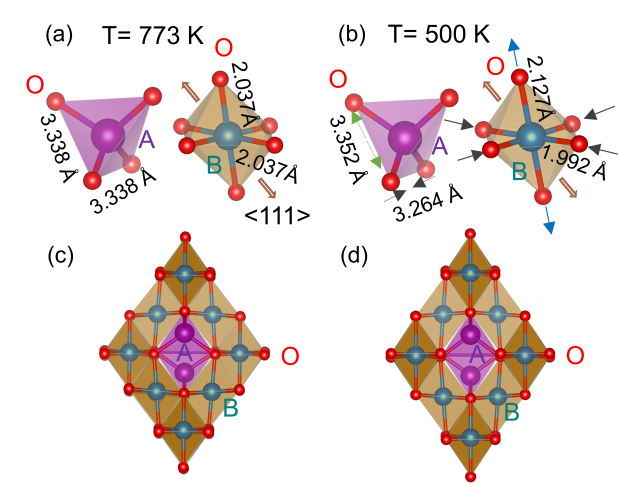}
\caption{Geometrical representation of the AO$_{4}$ (A=$Mn^{2+}$/$Fe^{3+}$) tetrahedron  and BO$_{6}$ (B=$Mn^{3+}$, $Fe^{3+}$ or $Mn^{2+}$) octahedron, and their projections in the ($HHL$) plane at 700 K(a,c)  and 500 K(b.d) . Thick open arrows in (a-b) illustrate the trigonal distortion of one octahedron at 773 K. Thin arrows in (a-b) indicate the evolution of the bond lengths in the tetrahedron and octahedron at 500 K.}
\label{fig:3}
\end{figure}

Figure~\ref{fig:3}(a-d) shows the geometrical representation of of the AO$_{4}$ (A=$Mn^{2+}$/$Fe^{3+}$) tetrahedron and BO$_{6}$ (B=$Mn^{3+}$/$Fe^{3+}$/$Mn^{2+}$) octahedron, and their projections in the ($HHL$) plane in the cubic notation. In the cubic structure at 773 K, all O-A-O bond angles
in AO$_{4}$ are the same $\approx$109.47$^{\rm o}$, forming a regular tetrahedron. All O-O bond lengths within the tetrahedron are the same. For the BO$_{6}$ octahedron, the O-B-O angle deviates significantly from 90$^{\rm o}$, with one being 83.66$^{\rm o}$ and another 96.34$^{\rm o}$. This can also be seen from the projection of the BO$_{6}$ octahedron in Fig.~\ref{fig:3}(c), reflected by a noncollinear ...O-B-O-B-O... chain due to the stretching of the octahedron in the $<111>$ direction.
 Such trigonal distortion is closely associated with the oxygen position within the cubic  $F d -3 m$ space group. Our results indicate that the trigonal distortion is already present in the cubic phase of FeMn$_{2}$O$_{4}$. The angular separation is $\approx$12.6$^{\rm o}$ at 773 K is comparable to MnV$_{2}$O$_{4}$ (12.9$^{\rm o}$) \cite{Sarkar2009} and FeV$_{2}$O$_{4}$ (10.7$^{\rm o}$)\cite{MacDougall2012}. 
Trigonal distortion only modifies the O-B-O bond angles, but does not change the O-B bond lengths (2.037 \AA{} at 773 K). Note that the trigonal distortion extends to $T<T_{S}$ as shown in Fig.~\ref{fig:3}(b) and (d), indicating that this is an intrinsic structural feature of FeMn$_{2}$O$_{4}$.

\subsection{Magnetic structure determination in $T_{N2}<T<T_{N1}$ and $T<T_{N2}$}

\begin{figure}
\centering
\includegraphics[width=1\linewidth]{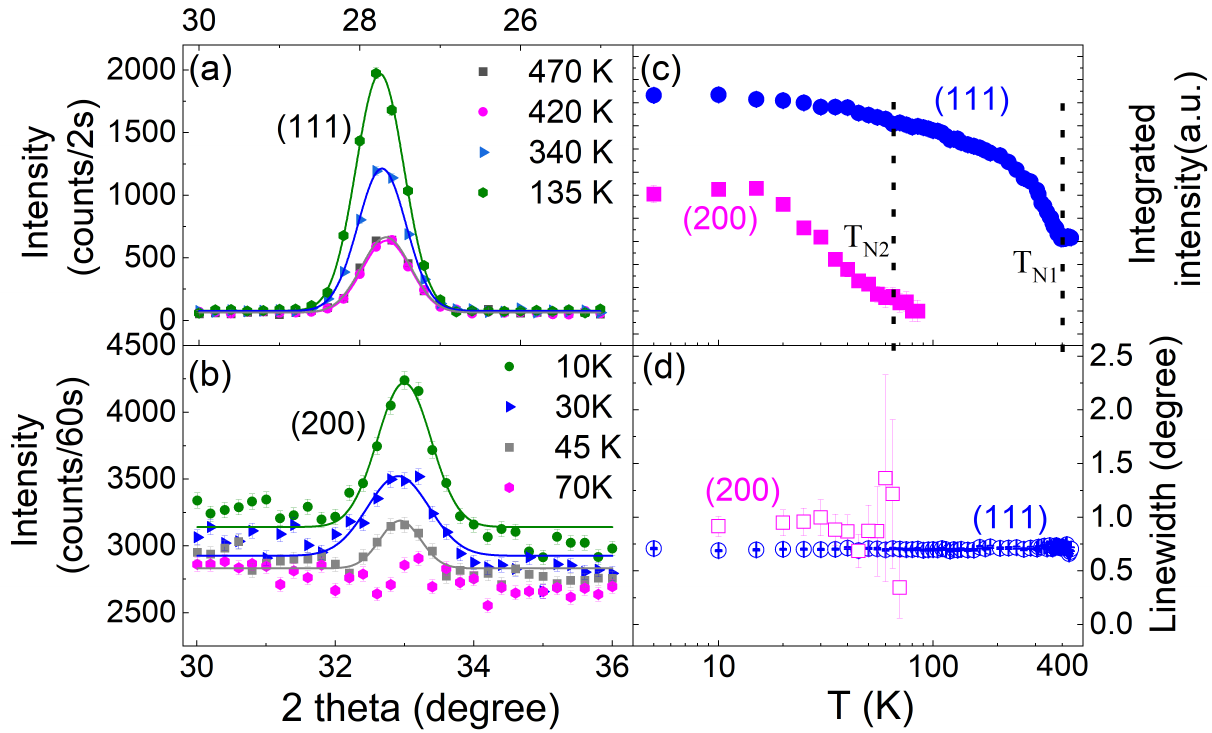}
\caption{ (a) Two theta scans of (a) (111) and (b) (200) peaks at indicated temperatures. The solid curves are the fits using a Gaussian function. Temperature dependence of the (c) integrated intensity and (d) linewidth of these two peaks.  
   }
\label{fig:4}
\end{figure}

 To identify the magnetic transition temperatures, the $2\theta$ scans of the (111) and (200) magnetic peaks at a number of temperatures were measured at the constant-wavelength triple-axis spectrometer HB1A. The representative results at indicated temperatures are shown in Fig.~\ref{fig:4}(a-b). 
The temperature dependence of the integrated intensity and linewidth of these two peaks are displayed in Fig.\ref{fig:4}(c-d), respectively. The rapid increase of the integrated intensity of the (111) peak below $T_{N1}\approx$ 400 K indicates a magnetic transition. The pure magnetic peak (200) emerges below the second magnetic transition $T_{N2}\approx$ 65 K.
Both $T_{N1}$ and $T_{N2}$ are consistent with those determined from the increase and decrease of magnetization, respectively\cite{Nepal2018}. Note that slightly lower $T_{N1}$  and $T_{N2}$  in our previous report\cite{Nepal2018} result from a different definition by using the peaks in the derivatives of the magnetization. Below $T_{N1}$ and $T_{N2}$, the magnetic peaks (111) and (200) are resolution limited indicative of long-range magnetic orders in $T_{N2}<T<T_{N1}$ and $T<T_{N2}$. There is no obvious increase of the linewidth for the (111) peak above $T_{N1}$ indicating that there is no short-range magnetic ordering. %
 Note that in addition to the (111) and (200) peaks, we have observed magnetic contribution at many nuclear peak positions such as (202), (220), (222), (313), (331). All the magnetic reflections can be indexed on the unit cell in these two temperature regions, and therefore, magnetic propagation vector is $\textbf{\textit{k}}=$(0,0,0).
 \begin{figure}
\centering
\includegraphics[width=1\linewidth]{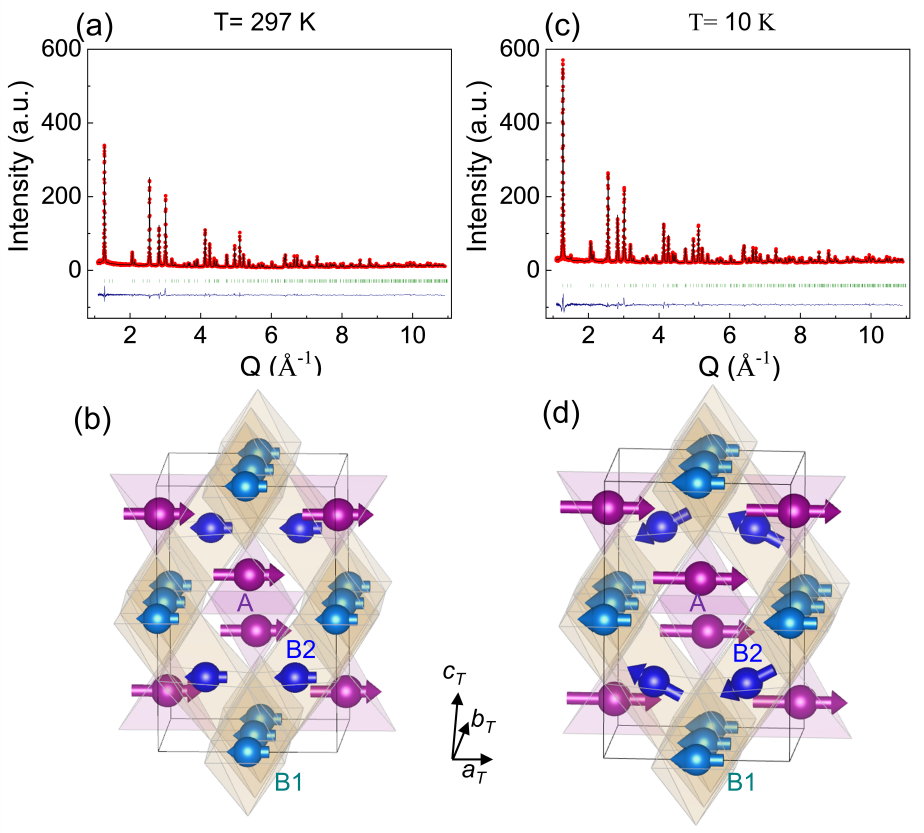}
\caption{ Rietveld refinement fits to high-resolution neutron diffraction patterns at (a) 297 K and (b) 10 K and the corresponding magnetic structures in (c) and (d). 
   }
\label{fig:5}
\end{figure}

To determine the magnetic structures, we analyzed symmetry-allowed maximal magnetic space groups and magnetic subgroups using the Bilbao Crystallographic Server\cite{Ivantchev:ks0038} and ISODISTORT\cite{Campbell2006} firstly. Then, we tested them one by one by fitting to the high-resolution POWGEN data over a large $Q$ region. Rietveld analysis of the POWGEN data, shown in Fig.~\ref{fig:5}(a) in  $T_{N2}<T<T_{N1}$ reveals a collinear FI order with the ordered moment along the tetragonal $a_{T}$ axis, as illustrated in Fig.~\ref{fig:5}(b). The magnetic space group is $Imm^{\prime}$a$^{\prime}$ (No.74.559). For this magnetic space group, there are two magnetic sublattices at the B site (namely B1(0 0 0.5) and the B2 (0.25 0.75 0.75)), in addition to one magnetic sublattice at the A site. The ordered moments at the B1 and B2 sites are found to be the same in this temperature region although the magnetic symmetry allows different moments. The moment at B1/B2 sites is antiparallel to that at the A site with a smaller magnetic moment, thus forming a
 $N\acute{e}el$-type collinear FI order. At 297 K, the ordered moment of the $(Mn^{2+}$/$Fe^{3+})_{A}$ site is (3.08(4),0,0) $\mu_{B}$, whereas the ordered moment at the B site ($Mn^{3+}$, $Fe^{3+}$ and $Mn^{2+}$) is (-1.61(3),0,0) $\mu_{B}$. The net ordered moment is (-0.14(5),0,0)$~\mu_{B}/f.u.$ ($3.08-2\times1.61=-0.14$) along the $a_{T}$ axis.

 In $T<T_{N2}$, a noncollinear magnetic structure involving a spin canting at the B2 site is found to best fit the neutron data. Rietveld refinement on the POWGEN data at 10 K is shown in Fig.~\ref{fig:5}(c). Compared to the magnetic structure in $T_{N2}<T<T_{N1}$, the main difference for $T<T_{N2}$ is that a moment component along the $+/-~c_{T}$ axis at the B2 site appears, leading to a spin canting in the $a_{T}c_{T}$ plane and a noncollinear FI order, as illustrated in Fig.~\ref{fig:5}(d). The projection of the canted spin along the $c_{T}$ axis shows an antiferromagnetic arrangement and is directly responsible for the emergence of the pure magnetic peak (200).  The moments at the A  site and the B1 site are constrained to point along the $a_{T}$ axis, whereas the moment at the B2 site is constrained in the $a_{T}c_{T}$ plane, as shown in Table III. At 10 K, the moments at the A and B1 sites are (4.65(5),0,0) and (-2.89(6),0,0) $\mu_{B}$, respectively. The moment at the B2 site is (-2.76(3),0,-1.31(7)) $\mu_{B}$, yielding a spin canting angle of 25$^{\rm{o}}$. The global net ordered moment at 10 K is $\approx$ (1,0,0) $\mu_{B}/f.u.$ along the $a_{T}$ axis. The structural and magnetic parameters at 10 K are listed in Table ~\ref{MagneticStructure}. 
 
  The magnetic structure we obtained here is distinct from that proposed in Ref.13 with magnetic space group $C2'/m'$ in both the magnetic moment size and directions. In Ref. 13, eleven nuclear and/or magnetic Bragg peaks were measured at a low-$Q$ region (Q$ <$ 3.3 \AA{}$^{-1}$) (see Fig. S2). Our attempt to refine these eleven Bragg peaks in our own data results in a few possible magnetic space groups that are consistent with the data for magnetic space groups $I4_{1}/am'd'$, $Imm^{\prime}$a$^{\prime}$, and $C2'/m'$. Even for the same magnetic subgroup $C2'/m'$ owing to its low symmetry, a few magnetic structures with different moment size and directions including that proposed in Ref. 13 are not distinguishable from the fits to these eleven peaks. Due to large moments in FeMn$_{2}$O$_{4}$, a substantial magnetic signal is observed up to $\approx$ 5.2 \AA{} (see Fig. SII). It turns out that adding the magnetic peaks in the $Q$ coverage 3.3 \AA{}$^{-1}$<Q<5.2 \AA{}$^{-1}$ are crucial to distinguishing different noncollinear ferrimagnetic orders. The high $Q$ data $>$5.2 \AA{}$^{-1}$ is also important to obtain the reliable structural parameters, such as the scale factor, thermal parameters, occupancy, and site inversion to separate the nuclear and magnetic contributions to the same peak and narrowing down the magnetic models. We find that the magnetic space group $Imm^{\prime}$a$^{\prime}$ best fits our POWGEN data. Interestingly, an attempt to use the magnetic space group $C2'/m'$ to refine POWGEN data including a larger number of magnetic and nuclear peaks than those in Ref.13 finds the same magnetic structure based upon the magnetic space group $Imm^{\prime}$a$^{\prime}$ proposed here, which further validated our magnetic model. Our symmetry analysis reveals that these two magnetic space groups are related to interpret this. For the same parent space group $I4_{1}/amd$ and magnetic propagation vector $\textbf{\textit{k}}=$(0,0,0), $Imm^{\prime}$a$^{\prime}$ is the maximal magnetic space group, whereas $C2'/m'$ with lower symmetry is one subgroup of $Imm^{\prime}$a$^{\prime}$. To the best of our knowledge, the ground-state magnetic structure of FeMn$_{2}$O$_{4}$ determined here is unique among all the reported spinel oxides in that only half of the B-site spins are canted but other half of the B-site spins are not.
 
 It is worthwhile noting that the single crystallographic B-site that is indicated for $T_{N1}<T<T_{S}$ is divided into two magnetic sites B1 and B2 in $T<T_{N1}$ with the magnetic space group $Imm^{\prime}$a$^{\prime}$ derived from the orthorhombic space group $Imma$. However, our careful Rietveld analysis on the high-resolution POWGEN data did not detect orthorhombic distortions, either in the lattice constants or  (B$_{1}$)(O$\_$1)$_{6}$/(B$_{2}$)(O$\_$2)$_{6}$ octahedra. The average crystal structure in $T<T_{N1}$ can be still described by its parent tetragonal space group $I4_{1}/amd$, consistent with that determined from previous x-ray diffraction\cite{BRABERS1971}. However, it is likely that the local structure is $Imma$ and there are local distortions of O$\_$1 and O$\_$2 suggesting the different local (B$_{1}$)(O$\_$1)$_{6}$ and (B$_{2}$)(O$\_$2)$_{6}$ environments to reconcile two magnetic sites in this temperature region. Further experiments on the local crystal structure making use of scanning tunneling microscope (STM) or Synchrotron/neutron total scattering techniques would be required to check this supposition. Pair distribution function analysis\cite{Zhang2022} on the total scattering data would also help explore if there is difference in the local structures for the different magnetic cations that are randomly distributed at the same A or B sites.

 \begin{table*}

 \centering
\setlength{\abovecaptionskip}{10pt}%
\setlength{\belowcaptionskip}{10pt}%
\caption{Refined structural and magnetic parameters from modeling high-resolution powder neutron diffraction data of FeMn$_{2}$O$_{4}$ at 10 K. The magnetic space group is determined to be $Imm^{\prime}$a$^{\prime}$ (No. 74.559) with parent space group $I4_{1}/amd$ and lattice constants of $a_{T} =$ 5.8945(2)\AA{} and $c_{T}=$ 8.8855(3)\AA{}}
\renewcommand{\arraystretch}{0.8}
\scalebox{1}{
 
\begin{tabular}{c|c|c|c|c|c|c|c|c|c|c|c}
\hline\hline
 label &spin valence &  x & y &  z& multiplicity  &  $B$ &  symmetry constrains on M &  M$_{x}$  & M$_{y}$&  M$_{z}$ &  $\vert$M$\vert$\\
 \hline
Mn1 &  Mn$^{2+}$&    0&  0.75     &   0.125  & 4&  0.320(6) & m$_{x}$,0,0 & 4.65(5)& 0 & 0 &  4.65(5)\\
Fe1&  Fe$^{2+}$ &    0&  0.75     &   0.125   & 4 &   0.320(6) &  m$_{x}$,0,0  & 4.65(5)& 0 & 0 & 4.65(5) \\ 
Fe2\_1 & Fe$^{3+}$ &    0&  0     &   0.5  &   4  & 0.266(4)& m$_{x}$,0,0 & -2.89(6) & 0 & 0 & 2.89(6)\\
Fe2$\_$2 & Fe$^{3+}$&    0.25&  0.75     &   0.75   &  4 & 0.266(4)& m$_{x}$,0, m$_{y}$ &  -2.76(3)  & 0 & -1.31(7) & 3.05(9)\\
Mn2$\_$1 & Mn$^{2+}$&    0&  0     &   0.5  &   4  & 0.266(4)& m$_{x}$,0,0 & -2.89(6) & 0 & 0 & 2.89(6)\\
Mn2$\_$2 & Mn$^{2+}$&    0.25&  0.75     &   0.75   &  4 & 0.266(4)& m$_{x}$,0, m$_{y}$ &  -2.76(3)  & 0 & -1.31(7) & 3.05(9)\\
Mn3$\_$1 & Mn$^{3+}$&    0&  0     &   0.5  &   4  & 0.266(4)& m$_{x}$,0,0 & -2.89(6) & 0 & 0 & 2.89(6)\\
Mn3$\_$2 & Mn$^{3+}$&    0.25&  0.75     &   0.75   &  4 & 0.266(4)& m$_{x}$,0, m$_{y}$ &  -2.76(3)  & 0 & -1.31(7) & 3.05(9)\\

O$\_$1 & O$^{2-}$&  0  &0.474(4)  & 0.261(5)    &  8 & 0.642(7) &   &     &  &   &  \\
O$\_$2 & O$^{2-}$&   0.775(8) &  0.75     &    0.511(6)   & 8& 0.642(7)&   &    &  &  & \\
 \hline\hline  

\end{tabular}}
\label{MagneticStructure}
\end{table*}

\subsection{Correlation between polyhedral distortions and structural transition at $T_{S}$}

\begin{figure}
\centering
\includegraphics[width=1\linewidth]{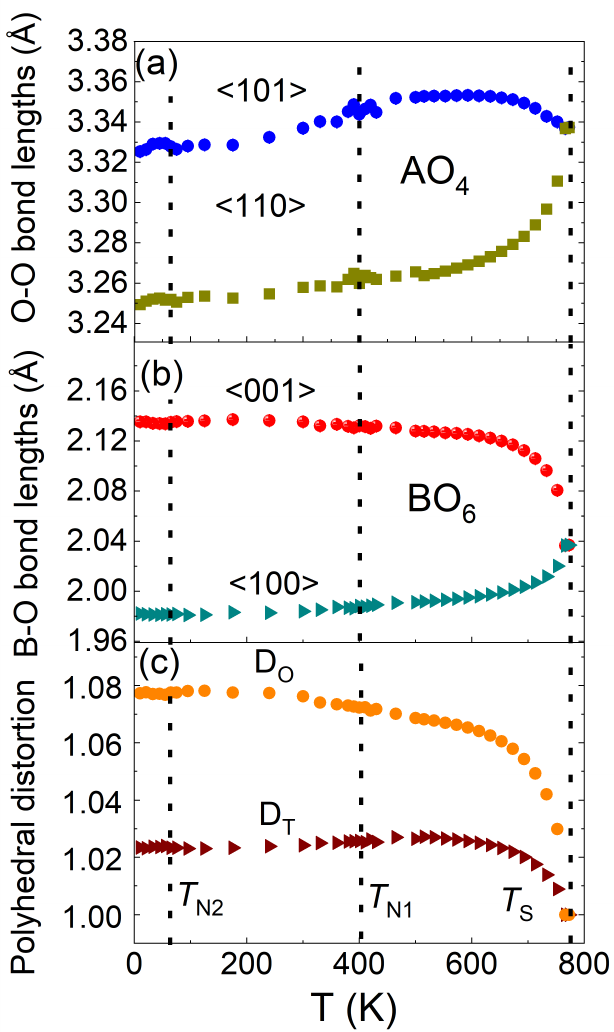}
\caption{ Temperature dependence of (a) the O-O bond lengths in tetrahedra AO$_{4}$ along  $<101>$ and $<110>$, and (b) the B-O bond lengths in octahedra BO$_{6}$ along $<001>$ and  $<100>$. (c). Temperature dependence of tetrahedral and octahedral distortion parameters D$_{\rm T}$ and D$_{\rm O}$.
   }
\label{fig:6}
\end{figure}

\begin{figure}  
\centering
\includegraphics[width=1\linewidth]{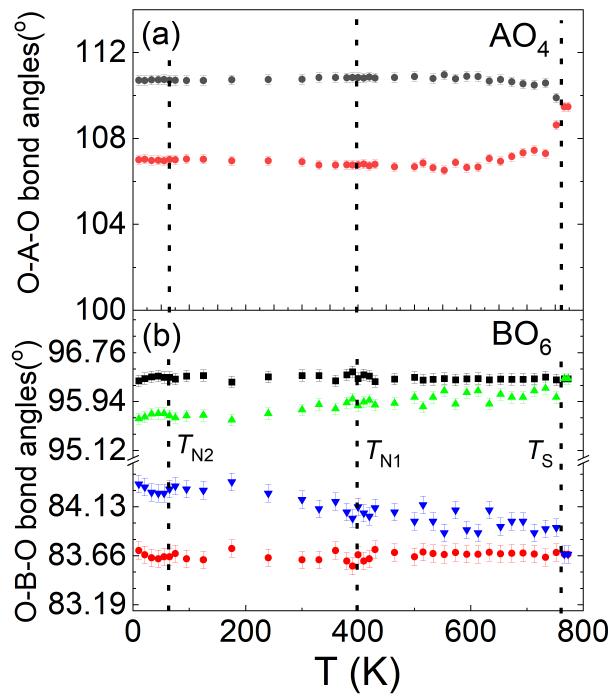}
\caption{Temperature dependence of (a) O-A-O bond angles in tetrahedral AO$_{4}$ and (b) O-B-O bond angles in octahedral BO$_{6}$.  
   }
\label{fig:7}
\end{figure}

To characterize polyhedral distortion, we examined the average bond lengths, angles and polyhedral distortion parameters based on Rietveld analysis of the neutron data. The distortion parameter of the AO$_{4}$ tetrahedron D$_{\rm T}$ is defined as the ratio of the O-O bond length along the $<101>$ direction to that along the $<110>$ direction, whereas the distortion parameter of the BO$_{6}$ octahedron D$_{\rm O}$ is characterized by the ratio of the B-O bond length along the $<001>$ direction to that along the $<100>$ direction.\cite{Katsufuji2008} The temperature dependence of the  bond lengths and angles in tetrahedra AO$_{4}$ and octahedra BO$_{6}$ is displayed in Fig.\ref{fig:6} and \ref{fig:7}, respectively. 
The temperature dependence of the distortion parameters D$_{\rm T}$ and D$_{\rm O}$ is shown in Fig. \ref{fig:6}(c). In the cubic phase, both D$_{\rm T}$ and D$_{\rm O}$ are 1 despite the existence of the trigonal distortion of the octahedron with split O-B-O bond angles. Upon cooling below $T_{S}$, the O-O bond length in the tetrahedron increases along the $<101>$ direction but decreases along the $<110>$ direction (see Figs.~\ref{fig:3}(b) and \ref{fig:6}(a)), leading to an elongated tetrahedron along the $c$ axis. Correspondingly, the tetrahedral distortion parameter D$_{\rm T}$ is found to increase from 1 at 773 K to 1.027 at 500 K. Below $T_{S}$, one O-A-O bond angle within the tetrahedron at 773 K splits into two, (110.88$^{\rm o}$ and 106.69$^{\rm o}$) at 500 K (see Fig. \ref{fig:7}(a)).

For the octahedron, the B-O bond length is elongated significantly along the $<001>$ direction but shortened along the $<100>$ direction (see Fig.~\ref{fig:3}(b) and \ref{fig:6}(b)), yielding a large D$_{\rm O}$ of 1.068 at 500 K. Below $T_{S}$, one O-B-O angle (96.34$^{\rm o}$ at 773 K) in the octahedron splits into two (96.35$^{\rm o}$ and 96.01$^{\rm o}$ at 500 K), whereas another O-B-O angle (83.66$^{\rm o}$ at 773 K) changes to 83.65$^{\rm o}$ and 83.99$^{\rm o}$ at 500 K (see Fig. \ref{fig:7}(b)). The elongated tetrahedron and 
octahedron along the $c$ axis lead to the cubic-tetragonal
structural transition with elongated lattice constant $c=c_{T}$ and shortened lattice constant $a=\surd2 a_{T}$ with $c > a$. The bond lengths, angles and the polyhedral distortion parameters at 773 and 500 K are also displayed in Table~\ref{crystal_1}, and~\ref{crystal_2}, respectively. Our results reveal a correlation between the cubic-tetragonal structural transition and the elongations of both tetrahedra and octahedra.

\subsection{Correlation between polyhedral distortion and magnetic order}
 
\begin{figure}
\centering
\includegraphics[width=1\linewidth]{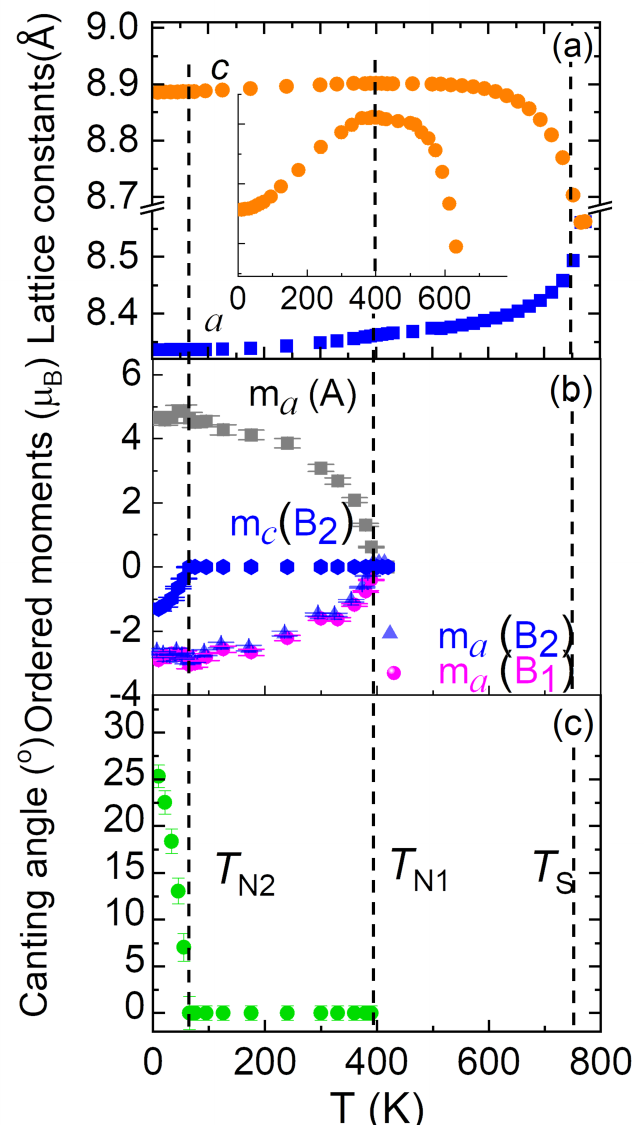}
\caption{Temperature dependence of (a) lattice constants in cubic notation, (b) the ordered moments at the A, B1 and B2 sites, and (c) spin canting angle at the B2 site (one half of the B site in pyrochlore lattice). The inset of the (a) shows the zoomed-in view for the lattice constant $c$.  
   }
\label{fig:8}
\end{figure}

 Fig.~\ref{fig:8}(a-c) shows the temperature dependence of the lattice constants, the ordered moments at different magnetic sites, and spin canting angle, respectively. Below $T_{N1}$, an anomalous decrease of the O-O bond length along the $<101>$ direction in tetrahedron AO$_{4}$ is observed while the
  O-O bond length along the $<110>$ direction continues to decrease (see Fig. 6(a)), which leads to a decrease of D$_{\rm T}$,  i.e., less tetrahedral elongation (see Fig.~\ref{fig:6}(c)). Note that compared to the ideal tetrahedron in the cubic phase in $T>T_{S}$, the tetrahedron is still elongated in this temperature region. On the other hand, we find an increase of the B-O bond length along the $<001>$ direction below $T_{N1}$, without clear anomaly in the B-O bond length along the $<100>$ direction (see Fig. 6(b)). This results in an increase of D$_{\rm O}$, i.e., further octahedral elongation below $T_{N1}$ (see Fig. ~\ref{fig:6}(c)). There are no clear anomalies in the O-A-O and O-B-O bond angles at $T_{N1}$ (see Fig. ~\ref{fig:7}). As shown in Fig.~\ref{fig:8}(b), the collinear FI order is established at $T_{N1}$ and the ordered moments along the $a_{\rm T}$ axis at both the A and B sites increase upon cooling. These results indicate that the formation of the collinear FI order drives both tetrahedral and octahedral distortions, mainly on the changes of the bond lengths rather than bond angles at $T_{N1}$. The tetrahedral and octahedral distortions lead to an anomalous decrease of the lattice constant $c$ and $a$ at $T_{N1}$. The concurrent collinear FI order, polyhedral distortions and lattice constants at $T_{N1}$ indicate a strong magnetoelastic coupling denoted magnetoelastic coupling I. 
  
  Upon further cooling below  $T_{N2}$, a collinear to noncollinear FI order occurs accompanied with the spin canting to the $c_{\rm T}$ axis at the B2 site.  At $T_{N2}$, a weak anomaly in the O-O bond lengths along both  the $<101>$ and $<110>$ directions is observed with a similar temperature dependence. Thus, no clear anomaly in  D$_{\rm T}$ can be seen at $T_{N2}$  (Fig. ~\ref{fig:6}(c)). There is no anomaly in the O-A-O bond angles in the tetrahedron at $T_{N2}$ either (see Fig. ~\ref{fig:7}(a)). Similarly, there is no obvious anomalies in the B-O bond lengths along the $<001>$, $<100>$, and D$_{\rm O}$. 
Interestingly, we find clear anomalies in O-B-O bond angles in the octahedron at $T_{N2}$ (see Fig.~\ref{fig:7}(b)), i.e., an anomalous decrease of the two low O-B-O angles ($\sim$83.6$^{\rm o}$ and 84.3$^{\rm o}$) with an increase of two high O-B-O angles $\sim$ 95.7$^{\rm o}$ and 96.4$^{\rm o}$. Therefore, the CFI-NCFI transition at $T_{N2}$ mainly induces changes of the O-B-O bond angles in the octahedron without a clear effect on the polyhedral distortion parameters, the O-A-O bond angles in the tetrahedron and the lattice constants, indicative of a different type of magnetoelastic coupling denoted magnetoelastic coupling II here.

\section{Discussion}
\subsection{Roles of the A-site Mn$^{2+}$, and B-site Fe$^{3+}$/Mn$^{3+}$ in the structure and magnetic properties}
   
Given that there is only about a 10 \% inversion between the A and B sites, we simplify the cation distribution of FeMn$_{2}$O$_{4}$ to ($Mn^{2+}$)$_{A}$($Mn^{3+}Fe^{3+})_{B}$O$_{4}$ to reveal the roles of the dominant A-site Mn$^{2+}$, and B-site Fe$^{3+}$/Mn$^{3+}$ in the structure and magnetic properties, by comparing with other related spinel oxides. 
    
\textit{Roles of B-site Fe$^{3+}$ cation} 

\begin{figure*}
\centering
\includegraphics[width=1\linewidth]{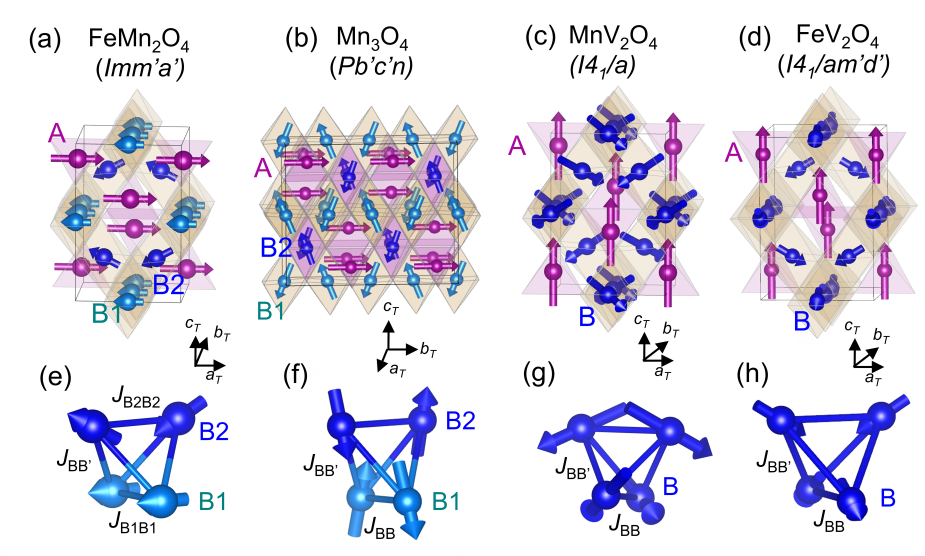}
\caption{ Comparison of the ground-state magnetic structures and magnetic space groups of (a) FeMn$_{2}$O$_{4}$, (b) Mn$_{3}$O$_{4}$, (c)  MnV$_{2}$O$_{4}$ and (d)  FeV$_{2}$O$_{4}$. The bottom panel shows the corresponding spin order in the B-site pyrochlore lattice only. Within the pyrochlore lattice, there are two magnetic sublattices B1 and B2 in FeMn$_{2}$O$_{4}$ and Mn$_{3}$O$_{4}$, whereas there is only one magnetic site in MnV$_{2}$O$_{4}$ and FeV$_{2}$O$_{4}$. The main magnetic interactions in the pyrochlore lattice are also labeled.
   }
\label{fig:9}
\end{figure*}

Compared to Mn$_{3}$O$_{4}$ which has a the cation distribution of ($Mn^{2+}$)$_{A}$($Mn^{3+}Mn^{3+})_{B}$O$_{4}$\cite{Jensen1974}, FeMn$_{2}$O$_{4}$ with the cation distribution  ($Mn^{2+}$)$_{A}$($Mn^{3+}Fe^{3+})_{B}$O$_{4}$) can be viewed as the resultant compound by replacing 1/2 of the B-site Mn$^{3+}$ by Fe$^{3+}$ in Mn$_{3}$O$_{4}$. Given that Fe$^{3+}$ ($t^{3}_{2g}e^{2}_{g}$) is not orbitally active, the existence of only 1/2 of the B-site orbitally active Mn$^{3+}$($t^{3}_{2g}e^{1}_{g}$) weakens the Jahn-Teller (JT) effect compared to Mn$_{3}$O$_{4}$, resulting in a significantly reduced cubic-tetragonal structural transition temperature from 1433 to 750 K. It also helps explain a smaller lattice distortion $c/a\approx$ 1.06 in FeMn$_{2}$O$_{4}$ compared to $c/a\approx$ 1.16 in Mn$_{3}$O$_{4}$\cite{Chung2008}. The magnetic properties change dramatically although both compounds belong to the same tetragonal space group below $T_{S}$. Mn$_{3}$O$_{4}$\cite{Jensen1974} exhibits three magnetic transitions.  A FI order occurs at $\approx$ 43 K, followed by an incommensurate spiral order below $\approx$ 39 K. Below $\approx$ 33 K, it shows a noncollinear ferrimagnetic order. As discussed above, FeMn$_{2}$O$_{4}$ shows two magnetic transitions. The ground-state magnetic structures are distinct between these two compounds, as illustrated in Fig.~\ref{fig:9}(a-b,e-f). In  Mn$_{3}$O$_{4}$ with the magnetic space group $Pb'c'n$ derived from orthorhombic space group $Pbcn$, there are two magnetic sublattices at the B site. While the moment of the A-site Mn$^{2+}$ points to the $b$ axis, the Mn$^{3+}$ spins at both the B1 (0 0 0.5) and B2 (0.25 0.75 0.75) sites are canted, with slightly different moment sizes (3.64 $\mu_{B}$ at the B1 site and 3.25 $\mu_{B}$ at the B2 site) and canting angles\cite{Jensen1974,Gallego2016}. The component on the $c$ axis for the B2-site Mn$^{3+}$ spins is aligned antiparallel, which leads to a magnetic unit cell doubled compared to the chemical unit cell, i.e., \textbf{\textit{k}}=(0,1/2,0). The spin ordering of the B-site Mn$^{3+}$ in the pyrochlore sublattice is displayed in Fig.~\ref{fig:9}(f). In FeMn$_{2}$O$_{4}$, the A-site Mn$^{2+}$ moment points to the $a_{T}$ axis, equivalent to the $b_{T}$ axis in Mn$_{3}$O$_{4}$ due to the tetragonal symmetry. In contrast, only the B2-site moment is canted to the $c_{T}$ axis of FeMn$_{2}$O$_{4}$, with the magnetic unit cell identical to the chemical unit cell. The B-site spin ordering in the pyrochlore sublattice of FeMn$_{2}$O$_{4}$ is shown in Fig. ~\ref{fig:9}(e).

 \textit{Roles of B-site Fe$^{3+}$/Mn$^{3+}$ cations }
 
 To understand the roles of the B-site cations,  we compare FeMn$_{2}$O$_{4}$ ((Mn$^{2+}$)$_{A}$(Mn$^{3+}$Fe$^{3+})_{B}$O$_{4}$) with MnV$_{2}$O$_{4}$ (Mn$^{2+}$)$_{A}$(V$^{3+})_{B}$O$_{4}$). MnV$_{2}$O$_{4}$\cite{Garlea2008} exhibits a collinear FI transition with net moment along the tetragonal $c_{T}$ axis near 60 K. At a slightly lower temperature, 58 K, a noncollinear FI order occurs with a net moment along the $c_{T}$ axis, accompanied by a simultaneous cubic-tetragonal structural transition to a space group $I4_{1}/a$. The structural transition is driven by  the orbitally active V$^{3+}$ ($t^{2}_{2g}e^{0}_{g}$) with the $t^{2}_{2g}$ degeneracy that induces a contraction of the VO$_{6}$ octahedron along the  $c_{T}$ axis yielding $c<a$. In contrast, Mn$^{3+}$ ($t^{3}_{2g}e^{1}_{g}$) involving the $e^{1}_{g}$ degeneracy in FeMn$_{2}$O$_{4}$ gives rise to the elongation of the octahedron along the $c_{T}$ axis with $c>a$. The distinct orbital degree of freedom between these two compounds results in different moment directions via spin-orbit coupling, i.e., along the $c_{T}$ axis in MnV$_{2}$O$_{4}$ and the $a_{T}$ axis in FeMn$_{2}$O$_{4}$. In addition, the noncollinear FI order of MnV$_{2}$O$_{4}$ below $T_{N2}$ is very different from FeMn$_{2}$O$_{4}$. The magnetic space group of MnV$_{2}$O$_{4}$ is $I4_{1}/a$ where there is only one magnetic B site with the moment canted to the same angle of 65.12$^{\rm o}$ relative to the A-site moment despite the B-site spin canting directions are different\cite{Garlea2008,Gallego2016} as illustrated in Fig. ~\ref{fig:9}(c) and (g).

 \textit{Roles of A-site Mn$^{2+}$ cations}

In Mn$_{3}$O$_{4}$, MnV$_{2}$O$_{4}$, and FeMn$_{2}$O$_{4}$, what is in common is Mn$^{2+}$ at the A site. The ordered moment for Mn$^{2+}$ in these three compounds is $\approx$ 4.5 $\mu_{B}$, indicating that it is in the high spin state ($t^{3}_{2g}e^{2}_{g}$). Thus, the A-site Mn$^{2+}$ is not an orbitally active cation and should not contribute to the structural transition in these compounds. However, the A-site Mn$^{2+}$ is necessary for forming both collinear FI and noncollinear FI orders via the antiferromagnetic interactions $J_{AB}$ with the B-site moment. The strength of $J_{AB}$ and the moment direction of Mn$^{2+}$ depend on the B-site cations. As characterized by the magnitude of $T_{N1}$, $J_{AB}$ is weak in Mn$_{3}$O$_{4}$ ($T_{N1}=43K$) and MnV$_{2}$O$_{4}$ ($T_{N1}=60K$), but is much stronger in  FeMn$_{2}$O$_{4}$ due to the presence of Fe$^{3+}$ at the B site. The moment direction of Mn$^{2+}$ points to the $c_{T}$ axis in MnV$_{2}$O$_{4}$, whereas it points to in-plane $a_{T}$ (or $b_{T}$) direction in Mn$_{3}$O$_{4}$ and FeMn$_{2}$O$_{4}$. As discussed above, this is related to the difference between $t^{2}_{2g}$ degenerate V$^{3+}$ and  $e^{1}_{g}$ degenerate Mn$^{3+}$. It should be noted that  there is a negligible change in the A-site Mn$^{2+}$ moment below/above $T_{N2}$ in all these three compounds since the magnetic transition mainly occurs in the B-site pyrochlore lattice. 
 
\textit{Important roles of cation distributions in spinel oxides}

At a first glance, the only difference between FeV$_{2}$O$_{4}$ and FeMn$_{2}$O$_{4}$ is V vs Mn. However, the structural and magnetic transitions of FeMn$_{2}$O$_{4}$ are very different from those of FeV$_{2}$O$_{4}$\cite{MacDougall2012}.  In FeV$_{2}$O$_{4}$, there exist three successive structural transitions: cubic-tetragonal ($c<a$) transition at $T_{S}=$138 K, a tetragonal-orthorhombic transition at  $T_{N1}=$ 111 K, and an orthorhombic-tetragonal ($c>a$) transition at $T_{N2}=$ 56 K. The latter two structural transitions are  accompanied with PM-CFI and CFI-NCFI transitions, respectively, with the net moments along the $c$ axis in each case. The differences in structures and magnetic transitions are due to completely different cation distributions in these two compounds, leading to distinct orbital and spin degrees of freedom. The cation distribution of  FeV$_{2}$O$_{4}$ is (Fe$^{2+}$)$_{A}$(V$^{3+})_{B}$O$_{4}$, in sharp contrast to (Mn$^{2+}$)$_{A}$(Mn$^{3+}$Fe$^{3+})_{B}$O$_{4}$ in FeMn$_{2}$O$_{4}$. There are two orbitally active cations Fe$^{2+}$ ($e^{3}_{g}t^{3}_{2g}$) at the A site and V$^{3+}$ ($t^{2}_{2g}e^{0}_{g}$) at the B site, responsible for the rich structural and magnetic transitions in FeV$_{2}$O$_{4}$. The JT distortion at the Fe$^{2+}$ site favors the compression of the tetrahedron, leading to cubic-tetragonal transition  with $c<a$. The cooperative orbital distortion manifested by an elongated FeO$_{4}$ and compressed VO$_{6}$ induces the low-T tetragonal structure with $c>a$. The orthorhombic structure in $T_{N2}<T<T_{N1}$ is the necessary intermediate regime between the switch of these two types of tetragonality. The net moments in both the CFI and NCFI regions point to the $c$ axis, with the magnetic space group $I4_{1}/am'd'$. In $T<T_{N2}$, all the V spins at the B site are canted, forming the ``two-in-two-out" ice-rule spin order in the pyrochlore lattice, with the same canting angle $\sim$ 55$^{\rm o}$ relative to the moment direction of the A-site Fe moment (see Fig. ~\ref{fig:9}(d) and (h). All of these results reinforce the importance of cation distribution to understand the crystal and magnetic structures of spinel oxides.

\subsection{Ordering processes and magnetic interactions}
  
\begin{figure*} 
\centering
\includegraphics[width=1\linewidth]{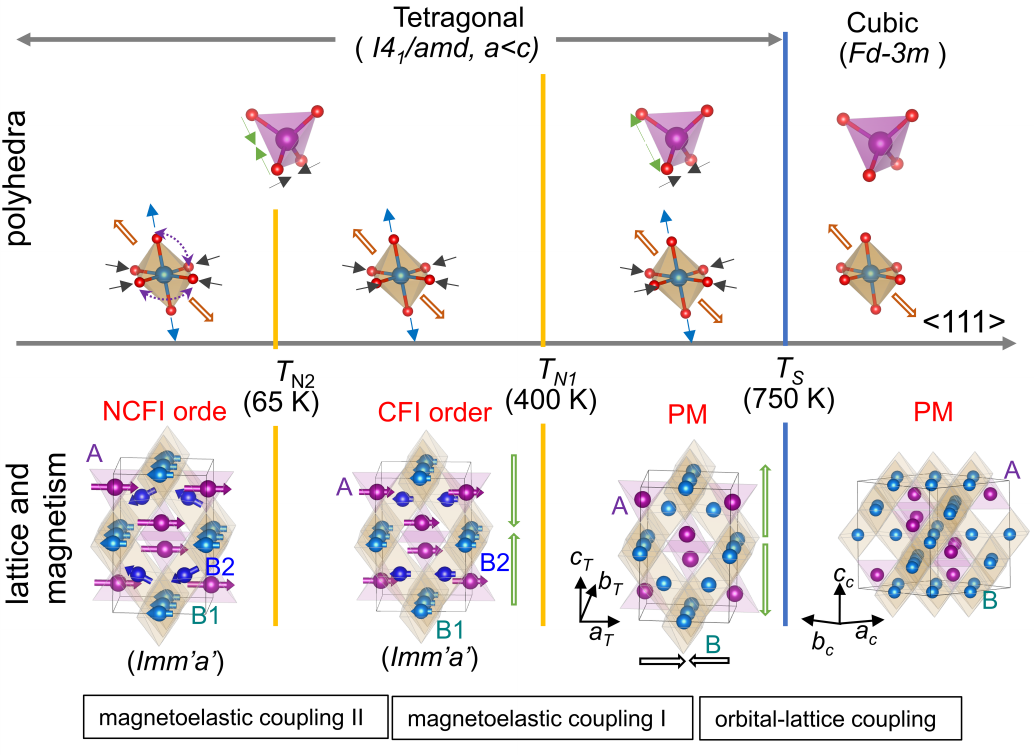}
\caption{ Polyhedral distortions, structural and magnetic ordering processes in FeMn$_{2}$O$_{4}$. Upon cooling, the new or modified arrows indicate the appearance of the new changes in the the tetrahedron or octahedron. In the top panel, the thick open arrows indicate the trigonal distortion along the $<111>$ direction and the thin arrows illustrate the evolution of the bond lengths in different temperature regions. The dashed purple arrows in $T<T_{N2}$ show representative and enlarged O-B-O bond angles near 95.7$^{\rm o}$ and 96.4$^{\rm o}$ (the correspondingly reduced O-B-O bond angles are not illustrated for clarity). The thick open arrows in the bottom panel show the change of the unit cell. Although the magnetic space group is $Imm'a'$ derived from orthorhombic space group in $T<T_{N1}$, the average crystal structure can be described by its parent space group tetragonal $I4_{1}/amd$.}
\label{fig:10}
\end{figure*}

 The polyhedral distortion, the structural and magnetic ordering processes in FeMn$_{2}$O$_{4}$ are summarized in Fig.~\ref{fig:10}. In the cubic structure ($F d -3 m$), while the AO$_{4}$ shows a perfect tetrahedron, the BO$_{6}$ shows large trigonal distortion along the $<111>$ direction. The effect of the trigonal distortion of VO$_{6}$ octahedra has been reported to vary in different vanadium spinels involving the $t^{2}_{2g}$ degeneracy. It has negligible effect on the degenerate t$_{2g}$ orbitals in ZnV$_{2}$O$_{4}$\cite{Maitra2007}. However, the trigonal distortion in MgV$_{2}$O$_{4}$ lifts the degenerate t$_{2g}$ orbitals and lowers the cubic symmetry from $F d -3 m$ to $F 4 -3 m$\cite{Wheeler2010}. In MnV$_{2}$O$_{4}$\cite{Chern2010}, trigonal distortion splits the t$_{2g}$ orbitals into a singlet and a doublet separated by an energy gap and reduces the site symmetry from cubic $O_{h}$ to $D_{3d}$. The effect of the trigonal distortion on orbitals in the cubic phase of FeMn$_{2}$O$_{4}$ is yet to be investigated.

 Upon cooling to $T_{S}$, the highest energy term that comes into play is the JT effect of orbitally active Mn$^{3+}$($t^{3}_{2g}e^{1}_{g}$) at the B site responsible for the cubic-tetragonal structural transition in FeMn$_{2}$O$_{4}$. The $e^{1}_{g}$ orbital degeneracy is lifted by the elongated octahedron along the $c$ axis, leading to the $d_{z^{2}}$ and $d_{x^{2}-y^{2}}$ singlets to host one electron. The distortion of oxygen atoms results in an elongation of tetrahedra along the $c$ axis due to corner sharing of oxygen atoms with octahedra. The elongated tetrahedra and octahedra along the $c$ axis result in cubic-tetragonal structural transition with  $c>a$ ($c/a\approx$ 1.06 at 500 K) in FeMn$_{2}$O$_{4}$, showing strong orbit-lattice coupling at $T_{S}$. Below $T_{S}$, both tetrahedral and octahedral distortion parameters D$_{\rm T}$ and D$_{\rm O}$ increase rapidly from 1 at 773 K to 1.027 and 1.068 at 500 K, respectively, indicating the enhancement of elongations of both the tetrahedron and octahedron. 
 
 As the temperature decreases to $T_{N1}$, the antiferromagnetic interaction $J_{AB}$ between the nearest-neighbor A and B spins, the second energy term, comes into play and leads to the antiparallel alignments of the A-site and B-site spins to form the $N\acute{e}el$-type ferrimagnetic order. Such a magnetic transition has strong effect on polyhedral distortion via spin-orbit coupling, yielding further octahedral elongation but reduced tetrahedral elongation below $T_{N1}$. The magnetic interaction $J_{AB}$ also induces a decrease of the lattice constant $c$ and a slope change in $a$ upon cooling, which is closely associated with its effect on the octahedral and  tetrahedral distortions. All of these results indicate a strong magnetoelastic coupling (magnetoelastic coupling I) at $T_{N1}$. 
      
Below $T_{N2}$, the important energy terms are the AFM interactions within the B-site moments in the pyrochlore lattice with geometrical frustration, which come into play and induces spin canting. Note that the A-site diamond lattice is not geometrically frustrated in FeV$_{2}$O$_{4}$ leading to the collinear moments at the A site. Spin canting transition induces modifications on the O-B-O bond angles by reducing two low O-B-O angles and enlarging other two high O-B-O angles in the octahedron, without significant changes in the the lattice constants and distortion parameters of tetrahedron/octahedron, revealing a distinct magnetoelastic coupling we label II. In Mn$_{3}$O$_{4}$\cite{Chung2008}, MnV$_{2}$O$_{4}$\cite{Chung2008}, and FeV$_{2}$O$_{4}$\cite{MacDougall2014}, two magnetic interactions within the pyrochlore lattice were found to be important: $J_{BB}$ in the $a_{T}b_{T}$ plane and $J_{BB'}$ out of the $a_{T}b_{T}$ plane as illustrated in Fig.~\ref{fig:9}(f-h). $J_{BB'}$ is weak in Mn$_{3}$O$_{4}$ and FeV$_{2}$O$_{4}$ due to the $c$-axis elongation of the BO$_{6}$ octahedra, but is strong and comparable to $J_{BB}$ in MnV$_{2}$O$_{4}$ due to the compressed BO$_{6}$ octahedra along the $c$ axis. The unique feature in the NCFI order of FeMn$_{2}$O$_{4}$ is that only the B2-site spins are canted while the B1-site moment direction remains unchanged. This suggests that besides $J_{BB'}$ out of the $a_{T}b_{T}$ plane, there are probably two types of $J_{BB}$ in the $a_{T}b_{T}$ plane, i.e.,  weak $J_{B1B1}$ and strong $J_{B2B2}$, as illustrated in Fig.~\ref{fig:9}(e). The competition of strong AFM $J_{B2B2}$ and AFM $J_{AB}$ induces a spin canting in the B2 site. But a small $J_{B1B1}$ cannot drive a spin canting within the B1 sublattice. Due to the $c$-axis elongation of BO$_{6}$ octahedra, $J_{B1B2}$ should be weak as overlapping of the neighboring orbitals of (Fe$^{3+}$/Mn$^{3+}$)$_{B}$ spins is expected to be weak. Our data suggest a somewhat complex hierarchy of magnetic exchange couplings in FeMn$_{2}$O$_{4}$. Further theoretical calculations and inelastic neutron scattering experiments are required to provide a quantitative measure of the magnetic exchange couplings and to explore the microscopic origin of the magnetic exchange constants such as different $J_{B1B1}$ and $J_{B2B2}$ in this compound.

 \section{Conclusions}

  In summary, we report the crystal and magnetic structures and two distinct magnetoelastic couplings in FeMn$_{2}$O$_{4}$. A first-order structural transition from cubic to tetragonal with large thermal hysteresis (120 K) is found at high temperature ($T_{S}\approx750$ K on warming). A large intrinsic trigonal distortion of the BO$_{6}$ octahedron exists even in the cubic phase and extends to the tetragonal phase, which however does not contribute to the structural transition. Instead, the structural transition results from the elongation of both tetrahedra and octahedra driven by orbitally active  Mn$^{3+}$ at the B site. This indicates the existence of an orbital-lattice coupling at $T_{S}$. Remarkably, we find the anomalies in polyhedral distortion and the lattice constants driven by the collinear FI order with the moment along the $a_{T}$ axis, reflecting a strong magnetoelastic coupling at $T_{N1}\approx400$ K. Such magnetoelastic coupling does not affect the bond angles in tetrahedra and octahedra. Below $T_{N2}\approx65$ K, a noncollinear FI order with spin canting at only half of the B sites, i.e., B2 site is found, which drives the anomalies in the O-B-O bond angles in octahedra without significant effect on the lattice constants and tetrahedral/octahedral distortion parameters characterized by the ratio of the bond lengths. Such a unique noncollinear magnetic ground state in the spinel family indicates that the magnetic couplings in the pyrochlore lattice can be distinct. We have also shown the importance of refining a wide-\textbf{$Q$} neutron data to separate the contribution from nuclear and magnetic contributions to the same peak(s) for solving the complicated $k=0$  magnetic order. The novel magnetic state and the interplay of spin, lattice and orbital degree of freedom presented here should motivate further experimental and theoretical work on FeMn$_{2}$O$_{4}$, its derivatives and more broadly other spinels.

\begin{acknowledgments}
Primary support for this study came from the U.S. Department of Energy under EPSCoR Grant No. DE-SC0012432, with additional support from the Louisiana Board of Regents. The neutron research used resources at High Flux Isotope Reactor and Spallation Neutron Source, DOE Office of Science User Facilities operated by the Oak Ridge National Laboratory.   \\

\textbf{Supporting information}\\
Rietveld refinement fits to high resolution neutron diffraction patterns at 500 K
after the thermal cycle and comparison of $Q$ regions between the neutron data in $Ref.$ 13 and our POWGEN data at 10 K.
 
\end{acknowledgments}

\bibliography{References_CM} 

\end{document}